\documentclass[12pt]{iopart}

\usepackage{graphicx}
\usepackage[caption=false]{subfig}
\usepackage{color}
\usepackage{amsfonts}

\newcommand{\bv}[1]{\mathbf{#1}}
\newcommand{\ket}[1]{\left \vert #1 \right \rangle}

\newcommand{\braket}[1]{\left \langle #1 \right \rangle}
\newcommand{\mrm}[1]{\mathrm{#1}}

\begin{document}

\title{Sensing phase-transitions via nitrogen-vacancy centers in diamond}

\author{P. Fern\'{a}ndez-Acebal and M.B. Plenio}
\address{Institut f\"{u}r Theoretische Physik and Center for Integrated Quantum Science and Technology (IQTS), Albert-Einstein Allee 11, Universitat Ulm, 89069 Ulm, Germany}

\date{\today}

\begin{abstract}
Ultra-thin layers of liquids on a surface behave differently from bulk liquids due to liquid-surface interactions. Some examples are significant changes in diffusion properties and the temperature at which the liquid-solid phase transition takes place. Indeed, molecular dynamics simulations suggest that thin layers of water on a diamond surface may remain solid even well above room temperature. However, because of the small volumes that are involved, it is exceedingly difficult to examine these phenomena experimentally with current technologies. In this context, shallow NV centers promise a highly sensitive tool for the investigation of magnetic signals emanating from liquids and solids that are deposited on the surface of a diamond. Moreover, NV centers are non-invasive sensors with extraordinary performance even at room-temperature. To that end, we present here a theoretical work, complemented with numerical evidence based on bosonization techniques, that predicts the measurable signal from a single NV center when interacting with large spin baths in different configurations. We therefore propose single NV centers as sensors capable to resolve structural water features at the nanoscale and even sensitive to phase transitions.
\end{abstract}

\maketitle

\section{Introduction}
Water is often referred to as one of the key molecules for life. Indeed, it takes a central role in many biological and chemical systems. Still, several properties of water remain under discussion; especially, the behavior of water at interfaces \cite{Bjorneholm2016,stevens2008simulations,hodgson2009water}. In particular, the hydrophilic behavior of biocompatible substrates has seen a significant increase in attention in recent years, with many studies reporting nucleation and ordering of water molecules on highly hydrophilic surfaces \cite{PhysRevLett.116.025501,algara2015square,chen2014unconventional}. Widely used methods of investigation, such as X-ray diffraction (XRD), neutron scattering or nuclear magnetic resonance (NMR), are limited by their spatial resolution and are highly affected by the thermal broadening. These disadvantages may be overcome using high-resolution techniques as atomic force microscopy (AFM), scanning tunneling microscope (STM) or similar \cite{guo2016perspective}. Still, the mentioned methods require low temperatures and high vacuum, making it hard to study water ordering at ambient conditions, which is the interesting regime for most applications, especially biology and medicine. In parallel, during the last decades new diamond production techniques have opened the route to novel applications in medicine and biology, making it a valuable candidate to substitute commonly-used titanium and stainless steel in medical prosthetics \cite{lappalainen1998diamond,lappalainen2005potential,dowling1997evaluation}. Yet, biocompatibility of diamond is poor since it may abrade internal tissues \cite{freitas2003nanomedicine}, and a biocompatible covering  such as water, may confer diamond the possibility to be used in such applications.

In this study we propose a sensing protocol based on shallowly implanted nitrogen-vacancy (NV) centers in bulk diamond that will help to detect the creation of thin water-ice layers covering the diamond surface. The designed technique is non-invasive, thus does not interfere with chemical processes occurring in water, and capable to determine the phase and behavior of water at ambient conditions. Our sensing tool consists on a NV center which possesses an electronic spin-1 in its ground state that can be easily polarized and read-out by optical means \cite{wu2016diamond}. The NV center electronic spin is stable and presents long coherence times at room-temperature even when shallowly implanted in bulk diamond \cite{jelezko2004observation,romach2015spectroscopy}. In fact, single nuclear spin sensing of immobilized spins has been demonstrated to be feasible within seconds \cite{muller2014nuclear}, while other studies suggest single nuclear spins can be detected in a similar time-scale even when undergoing random thermal motion \cite{bruderer2015sensing}. Furthermore, NMR measurements with sufficient resolution to resolve chemical shift have recently been achieved \cite{schmitt2017submillihertz,boss2017quantum,Du195}. Consequently, the NV center is an ideal candidate as a tool for detection and sensing of dense nuclear baths, as has been shown using decoherence measurement of electronic spins \cite{mcguinness2013ambient,staudacher2013nuclear}.

Here we do not only examine spin decoherence produced by a given bath, but the direct population exchange between the NV center and the surrounding nuclei in the Hartmann-Hahn double resonance (HHDR) regime \cite{hartmann1962nuclear}. Within this regime, the NV center is first driven with a microwave field, inducing Rabi oscillations on its electronic spin. Polarization transfer occurs  when the Rabi frequency is set to equal the energy splitting of certain nuclear spins. In this scheme, the interaction with a given set of nuclei is enhanced while possible interactions with undesired nuclear species, that is, those with different energy splitting, are suppressed \cite{cai2013,london2013detecting}. The HHDR performance dramatically depends on the nuclear motion, which is determined by the water phase. Therefore, measuring the NV center population indicates whether water is liquid or solid.

This paper is organized as follows: First, our theoretical set-up is described. We outline the required substrate for ice to be stable and introduce the NV center as a sensor for different water phases. Second, we develop a theory that predicts the measurable signal from an NV center spin interacting with the three plausible baths either liquid, solid or a mixture of both . Our approach combines a purely analytical prediction for fast diffusing spin with a bosonization protocol for the simulation of stationary nuclei, the latter of which is helping us to obtain numerical predictions. Finally, both numerical and theoretical results are presented and compared, suggesting that phase differentiation is possible in a short time-scale at room-temperature even for ultra-thin water layers. Moreover, we show that applying a magnetic field gradient along the quantization axis of the NV center we can obtain a faithful description of the crystal structure of solid ice.

\section{System description} 

\begin{figure}
	\includegraphics[width=1\columnwidth]{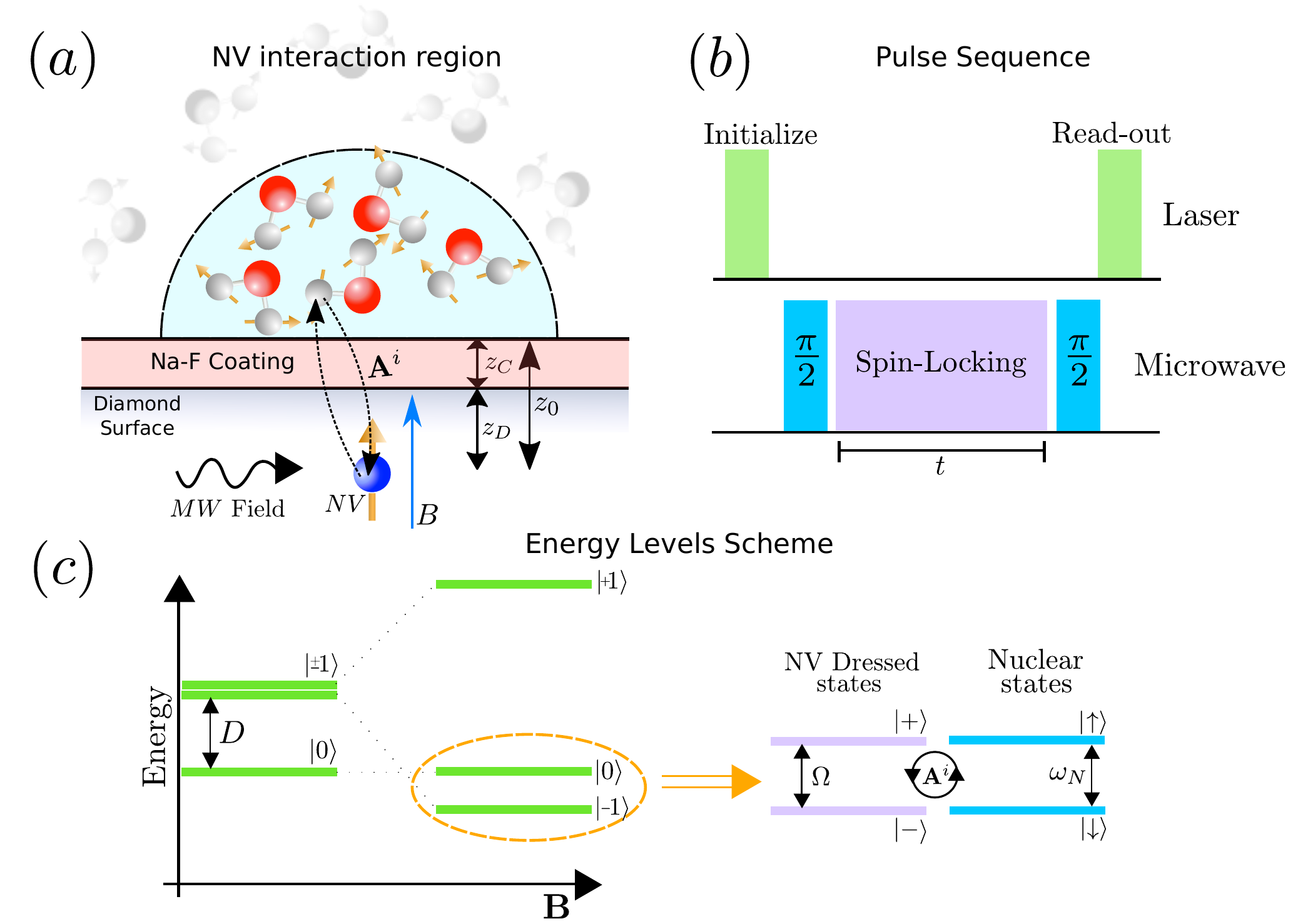}
	\caption{Schematic set-up for water phase sensing with shallow NV centers in bulk diamond and important parameters. (a) The NV center is located at a distant $z_0$ beneath the diamond surface. The water molecules are deposited on top of the diamond. The $^1H$ spins conforming the water move stochastically when liquid, stay fix when solid and may stay fixed in few layers while diffusing on top when the two phases coexists. The NV center interacts via dipole-dipole magnetic interaction, with a strength $g_i$. (b) Pulse sequence applied to measure the NV polarization loss. The NV is initially polarized and read-out using green-laser. During the microwave driving, $t$, the NV interacts with the nuclear spins at the HHDR condition. (c) Energy levels of the NV center and nuclear system. In the absence of magnetic field the NV is a spin$-1$ systems with projections $m_s = 0, \pm 1$, even at zero-field it exist a splitting of $D=2.87 \, \mrm{GHz}$. Degeneracy between $m_s=\pm 1$ levels is lifted applying an external magnetic filed, $B$, parallel to the NV axis. The transition between $m_s= 0$ and $m_s=-1$ is driven with microwave field inducing a Rabi frequency $ \Omega$ which matches the nuclear Larmor frequency $\omega_N = \gamma_N B$ permitting polarization transfer. 
	}
	\label{fig:Fig1}
\end{figure}

According to \cite{Wissner-Gross2007,Meng2006}, water ice may be stable on a $(111)$ diamond surface even above room temperature, since this surface has minimal mismatch with ice Ih crystal lattice. In fact, this lattice similarity reduces the structural strain, thus helping ice to stabilize. This effect is inherent to the $(111)$ surface and is not expected for $(100)$ nor $(110)$. Yet, strain reduction alone does not ensure a proper stabilization \cite{conrad2005ice}, indeed diamond is hydrophobic. In light of the results reported in \cite{Meng2006}, the hydrophilic behavior is increased when the surface is chemically modified using atoms with high water affinity. Hence, we presume here an alkali based coating where $1/3$ of diamond terminations are synthetically substituted by Na, while the remaining terminations are saturated by F atoms. Other coatings have been numerically proven to give rise to higher stabilization such as Na-H, which is expected to help to stabilize several ice layers. However, in this model hydrogen-based coatings are not considered since they interfere with the signal emanating from water protons. Also, even in the absence of H, the coating itself may adversely affect the charge state of the NV center. Throughout this work, these effects are not considered and it is assumed that the coating does not play a major role for the NV dynamics.

The sensing protocol is performed using a single shallowly implanted NV center in bulk diamond, Fig.(\ref{fig:Fig1}). The NV is assumed to be placed at a distant $z_0 \approx 3 \, {\mrm{nm}}$ beneath the coating surface. Water is deposited on top of the chemically modified diamond at a temperature that ensures ice formation. Then, the system is heated up to room temperature, which is our regime of interest, performing afterwards the experimental measurements. The polarization loss of the NV center is straightforwardly measured using a spin-locking sequence, Fig.(\ref{fig:Fig1}.b). First, the NV is optically polarized using $ 532 \, \mrm{nm}$ laser light. Afterwards, a $\pi/2$ pulse is applied and a suitable microwave field is used in order to provide a spin-locking fulfilling the HHDR condition. During this period the NV exchanges its polarization with a selected nuclear species and its environment. Finally, a second $\pi/2$ pulse rotates back the NV center spin and its state is measured by fluorescence spectroscopy.

In the following we consider the measurable signal from the NV center emanating from three plausible scenarios. First, the water becomes liquid at room-temperature, suggesting the coating is hydrophobic and does not serve as ice-substrate. Second, the water remains frozen, indicating the hydrophilic behavior of the surface. Third and more realistic, only few layers of ice are stabilized on top of the diamond while the remainder becomes liquid.

\section{System Hamiltonian}
Our system consists of a single NV center interacting with $N$ hydrogen spins-$1/2$ via dipole-dipole magnetic interaction. The NV center possesses an electronic spin-$1$ with projections $m_s= 0, \pm 1$. Applying an external magnetic field, $\bv{B}$, parallel to the NV quantization axis the degeneracy between $m_s = \pm 1$ levels is lifted. Additionally, using a MW field resonant with the $m_s = 0 \rightarrow m_s = -1$ transition allows us to consider the NV as an effective qubit. Furthermore, the external field, $\bv{B}$, induces a Larmor frequency on the nuclear spins $\omega_N = \gamma_N \vert \bv{B} \vert $, where $\gamma_N$ is the $^{1} H$ gyromagnetic ratio. In the dressed state basis for the NV, $  \ket{\pm} = \frac{1}{\sqrt{2}} \left( \ket{0} \pm \ket{-1} \right)$, the Hamiltonian of the system  in the secular approximation with respect to the NV reads ($\hbar=1$) \cite{london2013detecting,fernandez2017oil}

\begin{equation}
\label{eq:HamiltonianLiquid}
H= \Omega S_z + \sum_{i=1}^N \omega_N I_z^i + \left( S_x - \frac{1}{2} \right) \sum_{i=1}^N \bv{A}^i (\bv{r}_i) \bv{I}^i + H_{N-N},
\end{equation}
where $\Omega$ is NV center Rabi frequency, $ \bv{S}$ is the spin-$1/2$ operator in the dressed state basis, $\bv{I}^i$ is the spin operator of the $i-\mrm{th}$ nucleus and $\bv{A}^i(t)$ is the hyperfine vector between the NV center and the $i-\mrm{th}$ nuclear spin, which depends on the relative position between NV and nuclei, $\bv{r}_i$, and $H_{N,N}$ is the internuclear coupling.

\subsection{First scenario: Liquid water}
We first assume that water remains liquid at room temperature, corresponding to water on hydrophobic substrates. In this scenario, nuclear motion is fast, hence internuclear interactions represented by $H_{N-N}$ are averaged out to become negligible. Besides, nuclear random motion leads to a time-depending stochastic coupling, $\bv{A}^i(\bv{r}_i(t))$, with certain correlation time $\tau_c$, which depends on the water diffusion coefficient and the NV depth as  $ \tau_c \propto z_0^2 / {\cal{D}}_{W}$ \cite{fernandez2017oil,PhysRevB.92.184420}.

For water at room temperature the diffusion coefficient is large, ${\cal{D}}_W \approx 2 \cdot 10^3 \, {\mrm{nm}}^2 \mu {\mrm{s}}^{-1} $, which in combination with a shallowly implanted NV center leads to, $\tau_c  \approx 2 \, \mrm{ns}$. Consequently, on time scales exceeding $\tau_c$, the system state is described by $\braket{\rho} (t) = \rho_{NV} (t) \otimes \rho_B$ \cite{fernandez2017oil}, where $\braket{\rho}(t)$ is the system density matrix averaged over all possible stochastic trajectories, $\rho_{NV}(t)$ is the NV center density matrix and $\rho_B$ describes the nuclear bath, which is thermal. In this time scale, is possible to derive a dynamical equation for the average NV population $n(t) \equiv \frac{1}{2} + {\mrm{Tr}} \left( S_z \rho_{NV}(t) \right)$,

\begin{equation}
\label{eq:EqLiquid}
\dot{n}(t) + \alpha(t) n(t) = \frac{1}{2} \alpha(t),
\end{equation}
where $\alpha(t) = N \frac{1}{4} \left( \gamma_x(\Delta,t) + \gamma_x(2 \, \Omega-\Delta,t) + \gamma_z (\Omega,t) \right) $, being $\Delta = \Omega- \omega_N$ the HHDR detuning, which vanishes in our set-up. A detailed derivation may be found in \ref{sec:AppendixA}. The rates are defined as 

\begin{equation}
\gamma_\beta(\omega,t) = \int_0^t \braket{A_\beta^i(\tau) A_\beta^i (0)} \cos \left( \omega \tau \right) d \tau.
\end{equation}
Notice that in a Markov approximation, possible when $t \gg \tau_c$, the upper limit of the integral may be extended to infinity. In this regime, the rates $\gamma_{\beta}(\omega,t)$ correspond to the power spectra of the fluctuations evaluated at different frequencies. Consequently, the Markov approximation implies a time-independent rate, $\alpha_M \equiv \alpha(t \rightarrow \infty)$, leading to the average population at time $t$

\begin{equation}
\label{eq:DepolarizationEq}
n(t) = \frac{1}{2} + \frac{1}{2} \exp(- \alpha_M t).
\end{equation}
As a result, the NV center looses its initial polarization at rate $\alpha_M$. We remark that this procedure is similar to that presented in \cite{fernandez2017oil}. Nonetheless, due to the rapid motion of the water molecules, there is no net polarization transfer to the nuclear bath. Yet, the  protocol here presented is suitable for nuclear sensing. We remark, the depolarization rate $\alpha_M$ depends on the number of spins inside the NV detection volume, $N$, the correlation time, $\tau_c$, and the variance $\braket{A_\beta^i (0) A_\beta^i (0)}$, which are extensive parameters that in turn depend solely on a set of intensive parameters as $\lbrace \rho_W, z_0, {\cal{D}}_W \rbrace$, with $ \rho_W$ the proton density in water, and $z_0$ the vertical distance from the NV to the sample. Those are measurable quantities, which ensure a fit-free model for the population evolution.

\subsection{Second Scenario: Solid water}
In the case that water remains solid on top of the coated diamond surface, as predicted in \cite{Meng2006,Wissner-Gross2007}, the system evolution is again ruled by Hamiltonian in Eq.(\ref{eq:HamiltonianLiquid}), but with time-independent interactions since all the involved spins, NV center and nuclei, are now essentially stationary. Moreover, the internuclear coupling, previously neglected due to the rapid motion, now plays a significant role. The nuclear-nuclear interaction, $H_{N-N}$ may be written as

\begin{equation}
H_{N-N} = \frac{1}{2} \sum_{i,j} f_{i,j}(\bv{r}_{i,j}) \left( \bv{I}^i \bv{I}^j - 3 \left( \bv{I}^i \hat{\bv{r}}_{i,j} \right) \left( \bv{I}^j \hat{\bv{r}}_{i,j} \right) \right),
\end{equation}
where $f_{i,j}$ is the dipole-dipole interaction strength, and $\bv{r}_{i,j}$ are the position vectors joining two different nuclei. The enormous number of spins within the detection volume makes the derivation of a close analytical solution for $n(t)$ challenging, and an exact numerical treatment impossible. In order to gain insight into dynamical behavior of the system, we make use of the Holstein-Primakoff approximation (HPA) \cite{PhysRev.58.1098}, which considers polarized spins as bosons, such that $ S^+ \rightarrow a^\dagger$. When using HPA, the resulting system is Gaussian and therefore its evolution can be efficiently computed employing the covariance matrix, $\gamma_{i,j} = \braket{ a_i^\dagger a_j}$ \cite{Weedbrook2012,PhysRevA.49.1567}. In fact, the HPA has been proven to accurately work for highly-polarized spins and it is expected to give satisfactory results even for a thermal spin bath \cite{PhysRevB.75.155324}. The bosonic Hamiltonian after a HPA and a rotating-wave-approximation reads

\begin{equation}
\label{eq:HamiltonianSolidWater}
H= \Omega a^\dagger_0 a_0 + \sum_{i=1}^N \omega_i a^\dagger_i a_i + \sum_{i=1}^N g_i a^\dagger_0 a_i + \frac{1}{8} \sum_{i,j} h_{i,j} a^\dagger_i a_j + {\mrm{h.c.}} ,
\end{equation}
with the effective nuclear Larmor frequency $\omega_i = \omega_N - \gamma_N \frac{1}{2} A^i_z$, the complex coupling constant $g_i = \frac{1}{4} \left( A_x^i -\mrm{i} A_y^i \right)$, and $h_{i,j}$ the internuclear coupling strength; the NV center is labeled with $i=0$. We remark that non-quadratic contributions have been neglected hence, obtaining a quadratic Hamiltonian characterized by flip-flop interactions among different spins. The evolution equation for the covariance matrix may be obtained easily from the Von-Neumann equation, $\dot{\rho} = - \mrm{i} \left[H, \rho\right]$, where $\rho$ is the density matrix of the system. Using that by definition $\gamma_{i,j} \equiv \mrm{Tr} \left( a_i^\dagger a_j \rho \right)$, it is found that

\begin{equation}
\label{eq:EqSolid}
\dot{\gamma} (t) = -\mrm{i} \left[ V ,\gamma(t) \right],
\end{equation}
with $V$ such that $H= \bv{R}^\dagger V \bv{R} $ being $ R= \left( a_0, \dots,a_N \right)^{T}$. Yet, the number of nuclei, $N$, remains the computational limiting factor. For that reason, only particles inside a given interaction volume are considered. Taking a hemisphere of radius $ 2 \, z_0$ around the NV center, the most relevant interaction is accounted for and we expect a faithful description of the real dynamics. The latter can be easily confirmed taking as a figure of merit the total interaction inside a volume of radius $R_{M}$, that is, $\Lambda (R_M) \equiv \sum_{ \vert \bv{r}_i \vert \le R_M} g_i (\bv{r}_i)$. In our system it is verified that $\Lambda(2 \, z_0) / \Lambda(R_M \rightarrow \infty) \approx 80 \, \%$.

\subsection{Coexistence of phases} 
A more realistic scenario assumes that only a few layers of ice are stabilized on the chemically modified diamond while the remainder remains liquid. In this situation, the NV center interacts with both, solid and liquid water, the latter diffusing freely above the stationary ice. An appropriate description is obtained when the two previous methods are combined. The solid phase is represented using HPA bosonization and evolves according to Eq.(\ref{eq:EqSolid}), while the liquid water depolarizes the NV center as predicted in Eq.(\ref{eq:EqLiquid}). The cumulative effect is described by

\begin{equation}
\label{eq:CovarianceMatrixMixedWater}
\dot{\gamma} = - \mrm{i} \left[ V, \gamma \right]  + \alpha(t) \left \lbrace \Delta , \gamma \right \rbrace + \alpha(t) \Delta,
\end{equation}
where $\Delta$ is a matrix with elements $\Delta_{i,j}=\delta_{i,0} \delta_{j,0}$, indicating that liquid water affects directly the NV center. The derivation in detail may be found in \ref{sec:AppendixB}. The depolarization caused on the ice nuclei by the moving water has been neglected since the internuclear interaction strength is several orders of magnitude smaller than the electron-nuclear coupling. 
Various works have studied the behavior of liquid water in the presence of ice \cite{Karim1988,hayward2001ice}. Two main differences with respect to water in bulk are found: molecular diffusion is slowed down within few nanometers from the ice layer and molecules tend to be ordered in the immediate proximity of ice. The effect of space-dependent diffusion may be computed via molecular dynamics simulations and modifies $\alpha(t)$ via its dependence on the correlation function. In conjunction with the polarization loss, the molecular ordering may be examined using NV magnetometry. 

\subsection{NV magnetometry}
In the presence of a magnetic field gradient parallel to the NV center quantization axis, the nuclear Larmor frequency depends on the relative position between NV and nuclei, $\omega_i = \gamma_H \bv{B} (z_i)$, where $z_i$ is the vertical distance from the NV center. Thus, when some layers of ice are formed all the $^{1} H$ in a given layer will precess with the same Larmor frequency, $\omega_i = \gamma_H \bv{B} \left( z_{Layer} \right)$. In this case, the Rabi frequency, $\Omega$, may be tuned to match the specific Larmor frequency of a desired layer. Providing a high magnetic gradient or a sufficiently deep NV, such that the detuning between the Larmor frequencies in adjacent layers is large compared to their coupling with the NV center, the NV will interact exclusively with a two-dimensional bath consisting of all the spins conforming a given layer. On the other hand, fast moving nuclei will depolarize the NV center in a broad region of frequencies. Moreover, in our set-up the effect of liquid water may be neglected in a short time scale compared with the depolarization rate, $ \tau \alpha_M \ll 1$. As a consequence, scanning different Rabi frequencies with the NV center, we can individually identify all layers that have remained solid. The theoretically calculated envelope of the resonant peaks may be estimated as

\begin{equation}
\label{eq:Envelope}
n(\Omega =\omega_i,z_{Layer}) = \frac{1}{2}+\frac{1}{2} \cos ^2 \left( \sqrt{\rho_{2D}^i}\frac{\beta}{z_{Layer}^2} \tau \right),
\end{equation}
where $\tau$ is a fixed interaction time, $\beta$ is a factor coming from the  dipolar strength and $\rho_{2D}^i$ is the surface density of protons in a given layer. Thus, we can estimate both the ice thickness and inter-layer distance. Also, changes of superficial density, $\rho_{2D}^i$, at the solid-liquid interface as predicted on $\cite{Karim1988,hayward2001ice}$ are measurable within our scheme.

\section{Results and Discussion}

\begin{figure*}[!htp]
	\includegraphics[width=1\columnwidth]{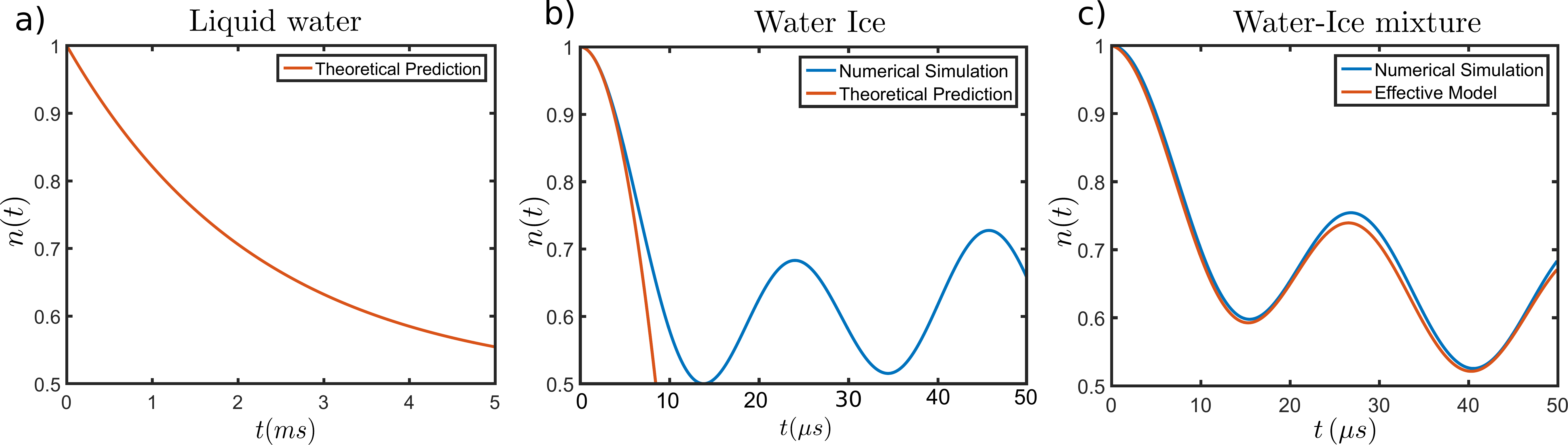}
	\caption{ Numerical simulation and theoretical prediction of the evolution for single shallow ($z_0 = 3 \, \mrm{nm}$)NV center polarization dynamics interacting with different baths. Note the different time-scales. (a) \textbf{Liquid water}. The NV center losses its polarization exponentially at a rate $\alpha^{-1}_M = 2.1 \, \mrm{ms}$, Eq.(\ref{eq:DepolarizationEq}). Notice that for rapid motion, in $\mrm{ms}$ time scale it is verified $t \gg \tau_c$. The depolarization is a consequence of flip-flop, flip-flip and pure dephasing processes, which are proportional to $\gamma_x(\Delta,t)$, $\gamma_x(2 \Omega - \Delta,t)$, and $\gamma_z(\Omega,t)$ respectively. (b)  \textbf{Water Ice}. The NV centers interacts with bulk ice governed by Eq.(\ref{eq:EqSolid}). For short-times an analytical prediction for a spin system is feasible, this gives a good approximation of the oscillations time-scale $\sum_i^N g^2_i$ (orange line). The HPA is used to obtain the behavior at longer time scales (blue line). Coherent oscillations are seen between the NV center and the protons in the ice. (c) \textbf{Water-Ice mixture}. The NV center interacts with a bath composed by a single ice bilayer and liquid water. The numerical simulation (blue line) was obtained neglecting the interaction with the liquid water. The effective model, (orange line) considers liquid water on top of the ice with space-dependent diffusion coefficient as calculated by \cite{Karim1988}, as can be seen liquid water has a negligible effect during this time-scale due to the fast motion.}
	\label{fig:Fig2}
\end{figure*}

The theoretical prediction made for liquid water may be compared with the numerical simulation performed with HPA for the last two cases. It must be recalled that, within this regime of parameters, a bosonization method for liquid water is proven to give inaccurate results (\ref{sec:AppendixB}), while an analytical formula for a solid bath does not exist.

The numerical and theoretical results are both depicted in Fig.~(\ref{fig:Fig2}). On the assumption that water is liquid, a shallow, $ z_0 = 3 \, \mrm{nm}$, NV center gets depolarized at a rate $\alpha_M^{-1} \approx 2.1 \, \mrm{ms}$, where we have assumed that water molecules will exhibit a diffusion constant similar to that in bulk, ${\cal{D}}_W = 2 \cdot 10^3 \mrm{nm}^2 \mu \mrm{s}^{-1}$. In fact, at room temperature water molecules diffuse rapidly interacting with the NV during a short time before diffusing away, so their net interaction is weak. In this regime, nuclear polarization is not possible. Indeed, for a net-polarization to happen an imbalance between flip-flop and flip-flip processes between NV and nuclei is required \cite{PhysRevB.92.184420}. The effect of flip-flop and flip-flip interactions is here represented by $ \gamma_x (\Delta,t)$ and $\gamma_x (2 \Omega - \Delta,t)$ respectively. For fast diffusion, meaning short correlation time, the power spectral density is flat in our frequency range making $ \gamma_x( \Delta,t) \approx \gamma_x (2 \Omega - \Delta,t)$. Hence, the nuclei do not gain polarization. Still, nuclear sensing is possible via NV center polarization loss.

On the other hand, when solid water is stabilized on the surface the measurable signal is marked by coherent oscillations in which the NV center interchanges its initial polarization with the surrounding bath. For short-times the polarization of a spin-system is proportional to $ n(t) \approx 1 - \frac{1}{2} \sum_{i}^N \vert g_i \vert^2 t^2 $, which can be evaluated to coincide with the HPA prediction. In effect, the time scale of the oscillations is determined by $\sum_{i=1}^N g_i^2 \approx 5 \, \mu s$. At larger times the oscillatory behavior persists, since for it to be suppressed the spectral density of nuclei in the ice phase must be dense around the Rabi frequency of the NV center, $\Omega$ \cite{van1992stochastic}, which is not true for the current system and thus, the polarization oscillates back and forth. For simulation purpose we have neglect the ice Ih residual entropy \cite{Thesis2010,pauling1988general,dimarzio1964residual}. In fact, we have considered the protons to be perfectly ordered inside the lattice since entropy effects on the lattice do not significantly affect the qualitative behavior of signal.

When the two phases coexists, a measurement at HHDR condition will lead again to coherent oscillations between the NV center and the solid layers. Liquid water depolarizes the NV center on a longer time-scale. According to \cite{Karim1988}, liquid water moves with a space-dependent diffusion coefficient at water-ice interface. It can be approximated as ${\cal{D}}_{W/I} (z) = {\cal{D}}_{min} + ({\cal{D}}_W - {\cal{D}}_{min}) \left[ 1+ \exp \left(-2 \kappa \left(z-z'\right) \right) \right]^{-1} $, where $\kappa$ and $z'$ are some parameters that may be obtained from \cite{Karim1988}. This behavior is easily integrated in our scheme since it only alters the correlation time $\tau_c$, which is straightforwardly calculated by numerical means. Computational results in this regime are included in Fig.(\ref{fig:Fig2}), revealing that even for a space-dependent ${\cal{D}}_{W/I}(z)$, the depolarization rate is small and thus effects from liquid water are negligible in our time scale. Hence, a thin layer of ice is detectable within few $\mu \mrm{s}$. 

\begin{figure}
	\includegraphics[clip,width=.5\columnwidth]{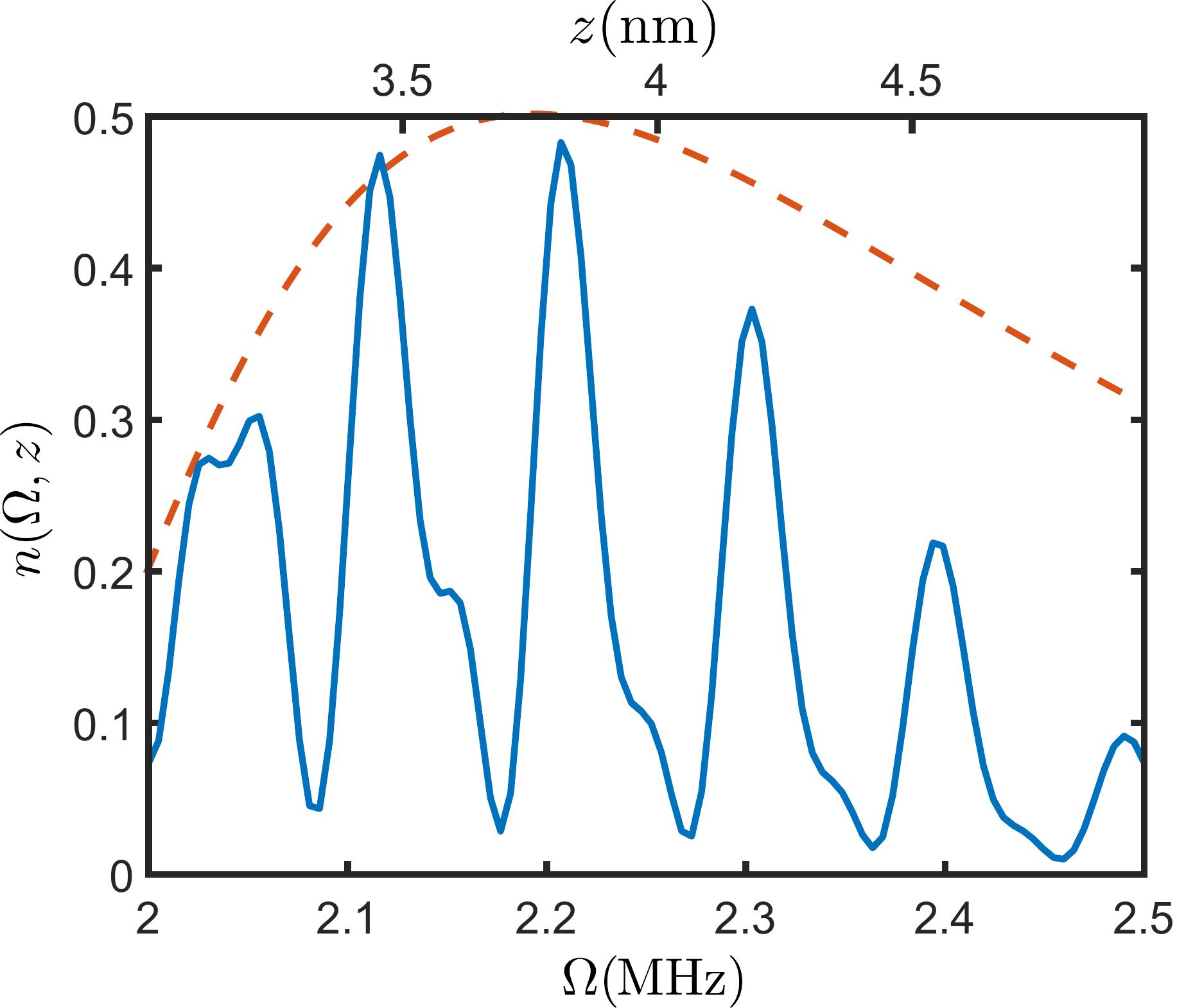}
	\caption{ Numerical simulation of NV center polarization exchange while interacting with bulk ice in the presence of a gradient field (blue line) compared with theoretical prediction of its envelope Eq.(\ref{eq:Envelope}) (orange line). Resonant peaks appear when the HHDR condition $\Omega = \omega_i = \omega_N + \frac{dB}{dz} z_i$ is fulfilled. At this condition, the NV centers interacts only with a given layer $i$, which results in a coherent polarization exchange with the bath at a frequency $\sum_j^{N_i} g_j^2$, where $j$ runs over the $N_i$ spins in $i$-th layer. The distance of each layer to the NV center is proportional to $\omega_i$ and can be easily determined from the measurement. For the theoretical have been used the estimated bidimensional density $\rho_{2D}^i = 15.2 \, \mrm{protons} \, \mrm{nm}^{-2}$ and $ \tau = 30 \, \mu \mrm{s}$. Deviations from the theory for larger distances corresponds to an artificial reduction of density due to numerical calculations. When the two phases coexist, only peaks corresponding to solid layers will be seen.
	}
	\label{fig:Fig3}
\end{figure}

Lastly, in the presence of a strong magnetic gradient, we can address individual solid layers. In fact, under a linear magnetic field gradient, $ dB/dz= 60 \, \mrm{G}/\mrm{nm}$, which has been already used in similar scenarios \cite{mamin2012high}, the resolution of the HHDR scheme is sufficient to discriminate different layers as it is seen in Fig.~(\ref{fig:Fig3}). Within this scheme structural properties such as the number of layers that remain solid, the interlayer distance or layer density fluctuations may be measured. Importantly, when a layer melts, water particles diffuse away rapidly thus the resonance peak disappear from the NV spectra for short times, conferring to this method high sensitivity. Thus, the number of observed peaks correspond to the number of stabilized layers.

Decoherence processes coming from external sources such as unpaired electron spins at the coating surface or paramagnetic impurities inside the diamond, can be neglected in our model. This is because in similar scenarios for driven shallowly implanted NV centers, the decoherence time has been measure to be $T_{1 \rho} \approx \mrm{ms}$ \cite{PhysRevLett.112.147602,PhysRevLett.110.017602}. As we are focused on a microsecond time range, $T_{1 \rho}$ relaxation does not need to be included.

\section{Conclusion}
In this work we have demonstrated that detection of phase transitions at room-temperature using shallow NV centers is feasible. More precisely, we have characterized the polarization loss produced on a single NV center by liquid, solid and liquid-solid large spin baths using HHDR dynamical decoupling. Our theoretical prediction for liquid water has been compared with extensive numerical analysis for solid ice. We have found that on a $\mu \mrm{s}$ time scale, NV center polarization undergoes coherent oscillations when interacting with solid water, while interactions with liquid water are negligible in the same time range. Even an ultra-thin ice layer of a few nanometers is detectable using such a protocol. These results permit unequivocal differentiation between solid and liquid water phases. Moreover, using a moderate-high magnetic field gradient, the thickness and density of the ice layer can be determined, being able to resolve nanometric structures.

Our approach may find applications in diverse frameworks. Definitely, it is extensible to other solid-liquid phase transitions involving magnetic nuclei, and since our protocol is sensible to density fluctuations, liquid-gas or solid-solid transitions are also detectable. In fact, transitions between different types of water ice, such as Ice Ih and Ice III, are also detectable using NV magnetometry. They are not included in this analysis since they are not expected to occur onto diamond. These results set the route to novel application of NV centers not only as phase transition sensors but also as valuable tools for nanostructural characterization.

\ack
We acknowledge fruitful discussions with Martin Bruderer and Nir Bar-Gill. This work was supported by the ERC Synergy Grant BioQ and the EU projects DIADEMS and HYPERDIAMOND.

\appendix

\section{Master equation for Liquid Water}
\label{sec:AppendixA}
In this Appendix we derive the master equation for the liquid water scenario that will lead to Eq.(\ref{eq:EqLiquid}). Starting from the Hamiltonian in Eq.(\ref{eq:HamiltonianLiquid}), and under the assumption that for fast-moving particles the internuclear interaction averaged out, we can rewrite the Hamiltonian as

\begin{equation}
\fl H = \Omega S_z + \sum_{i=1}^N \gamma_N \bv{B}_{\mrm{eff}}^i \bv{I}_i + \sum_{i=1}^N  g_i (t) S_+ I_-^i + g_i^*(t) S_+ I_+^i + h_i(t) S_+ I_z^i + \mrm{h.c.}
\end{equation}
In the last expression and for the sake of clarity, we have make use of the ladder operators to indicate the possible interactions. Also, we have renamed $g_i (t) \equiv \frac{1}{4} \left( A_x^i (t) + {\mrm{i}} A_y^i (t) \right)$ and $h_i(t) \equiv \frac{1}{2} A_z^i (t)$ and $\bv{B}_{\mrm{eff}}^i = \bv{B} - \frac{1}{\gamma_H}\frac{1}{2} \bv{A}^i \bv{I}^i$ is an effective field created on the nuclear spins by the NV center. First, we consider that we work in high-fields such that $ \gamma_N \vert \bv{B} \vert \gg \bv{A}_i \, \forall i$, and thus consider that all the nuclei rotate with identical Larmor frequency $\omega_N$. 

Second, in order to treat such a Hamiltonian one may split it into a time independent part, $H_\omega \equiv \Omega S_z + \sum \omega_N I_z^i$, and a time-dependent stochastic part

\begin{equation}
H_{int}(t) = \sum_{i=1}^N g_i (t) S_+ I_-^i + g_i^* (t) S_+ I_+^i + h_i (t) S_+ I_z^i + {\mrm{h.c.}}
\end{equation}
The system will evolve according to the Von-Neumann master equation,

\begin{equation}
\dot{\rho}(t) = - \mrm{i} \left[ H_\omega + H_{int} (t) , \rho(t) \right].
\end{equation}
This is a stochastic differential equation. Nonetheless, we want to look at the deterministic master equation, that is, averaged over all possible stochastic trajectories. There exist several methods for solving such an equation, in particular we follow \cite{van1992stochastic}, which have been used previously in \cite{bruderer2015sensing,fernandez2017oil}, obtaining accurate results. For solving the dynamics we go in a interaction picture with respect to $H_\omega$ , obtaining

\begin{equation}
\fl \tilde{H} =  e^{ {\mrm{i}} H_\omega t}  H_{int} (t)  e^{- {\mrm{i}} H_\omega t} = \sum_{i=1}^N  g_i (t) S_+ I_-^i e^{ \mrm{i} \Delta t} + g^*_i (t) S_+ I_+^i e^{ \mrm{i}(2 \, \Omega+ \Delta) t} + h_i(t) S_+ I_z^i e^{\mrm{i} \Omega t } + {\mrm{h.c.}},
\end{equation}
where we have introduced the detuning $\Delta = \Omega-\omega$. The Hamiltonian may be splitted in a time dependent and time independent part

\begin{eqnarray}
\fl \tilde{H}_0 = \braket{\tilde{H}(t)} = \sum_{i=1}^N  \braket{g} S_+ I_-^i e^{ \mrm{i} \Delta t} + \braket{g}^* S_+ I_+^i e^{ \mrm{i}(2 \, \Omega+ \Delta) t} + \braket{h} S_+ I_z^i e^{\mrm{i} \Omega t }  + {\mrm{h.c.}} , \\
\fl \tilde{H}_1 = \tilde{H}_0- \tilde{H}(t) = \sum_{i=1}^N {\cal{G}}_i (t) S_+ I_-^i e^{ \mrm{i} \Delta t} + {\cal{G}}^*_i (t) S_+ I_+^i e^{ \mrm{i}(2 \, \Omega+ \Delta) t} + {\cal{H}}_i (t) S_+ I_z^i e^{\mrm{i} \Omega t } + {\mrm{h.c.}}
\end{eqnarray}
where $\braket{\cdot}$ represent the average over all possible stochastic trajectories. We have introduced the hyperfine coupling fluctuations around the average ${\cal{G}}_i (t) \equiv g_i(t) - \braket{g}$, ${\cal{H}}_i (t) \equiv g_i(t) - \braket{H}$. Also, it is assumed all the particles posses identical average properties, $\braket{g_i} = \braket{g} \, \forall i$, $\braket{h_i} = \braket{h} \, \forall i$, which is true for homogeneous motion. We note that in the assumption that $\Delta \approx 0$, and $\Omega \gg  \Delta$, we can neglect fast rotating terms form $\tilde{H}_0$ obtaining

\begin{equation}
\tilde{H}_0 \approx \sum_{i=1}^N \braket{g} S_+ I_-^i + {\mrm{h.c.}} ,
\end{equation}
which is a time independent Hamiltonian where all the fast-rotating terms have been neglected. Moreover, in our scenario $\braket{g}= 0$.

Always in the interaction picture with respect to $H_\omega$, the master equation may be written as

\begin{equation}
\dot{\tilde{\rho}} (t) = - {\mrm{i}} \left[ \tilde{H} (t), \tilde{\rho}(t) \right] \equiv {\cal{L}}(t) \tilde{\rho}(t) = \left( {\cal{L}}_0 + {\cal L}_1 (t) \right) \tilde{\rho}(t).
\end{equation}
In the last expression we have made use of the Lioville superoperator, ${\cal{L}}_\alpha (t) \cdot \equiv - {\mrm{i}} \left[ \tilde{H}_\alpha, \cdot \right]$. Following straightforwardly reference \cite{van1992stochastic}, when $ \vert \tilde{H}_0 \vert \tau_c \ll 1$ and $\braket{{\cal{G}}^2} \tau_c \ll 1$, which is true for moderate coupling and short $\tau_c$, the average solution of such a system is

\begin{equation}
\dot{\braket{\rho}}(t) = - \mrm{i} \left[ \braket{H}, \braket{\rho}(t) \right] + \sum_{i=1}^N {\cal{H}}_i \braket{\rho}(t) + {\cal{D}}_i^x \braket{\rho}(t) +  {\cal{D}}_i^z \braket{\rho}(t),
\end{equation}
where $\braket{\rho}(t)$ is the average of the total density matrix $\rho(t)$ over all possible stochastic trajectories. This is an approximation up to second order in terms of $\braket{{\cal{G}}^n_i} \tau_c^{n-1}$. Up to this order, the effects of the stochastic motion can be split into three different contributions. A shift to the energy levels
\begin{equation}
\label{eq:MasterEqSpins}
{\cal{H}}_i \braket{\rho}(t) =  \mrm{i} \Omega_x(\Delta_i,t) \left[ S_z - I_z^i, \braket{\rho}(t) \right] + \mrm{i} \frac{1}{2} \Omega_z(\Omega,t) \left[S_z, \braket{\rho}(t) \right],
\end{equation}
with the frequencies defined as
\begin{equation}
\Omega_\beta \left(\omega,t\right) = \int_0^t \braket{A_\beta^i (\tau) A_\beta^i (0)} \sin \left( \omega \tau \right) d \tau.
\end{equation}
A dissipator originated from flip-flop and flip-flip interactions
\begin{eqnarray}
\fl {\cal{D}}_i^x \braket{\rho}(t) =  \gamma_x(\Delta_i,t) \left[ D \left( S_+ I_-^i \right) + D \left( S_- I_+^i \right) \right] \braket{\rho}(t) + \\ 
\gamma_x(2 \, \Omega + \Delta_i,t) \left[ D \left( S_+ I_+^i \right) + D \left( S_- I_-^i \right) \right] \braket{\rho}(t)
\end{eqnarray}
And the last term which will cause pure dephasing on the NV center
\begin{equation}
{\cal{D}}_i^z \braket{\rho}(t)  = \gamma_z(\Omega,t) \left[ D \left( S_+ I_z^i \right) + D \left( S_- I_z^i \right) \right] \braket{\rho}(t)
\end{equation}

We remark that for symmetry reasons, in our set-up it is verified $\braket{g} =0$. That means, there is not a net coherent coupling between NV center and nuclei, allowing us to express the system state as $\braket{\rho}(t) = \rho_{NV}(t) \otimes \rho_B$. The evolution of the NV population $ n(t) \equiv \frac{1}{2} + \mrm{Tr} \left( S_z \rho_{NV}(t) \right)$, may be calculated obtaining

\begin{eqnarray}
\label{eq:EvolutionEqSpinPopulation}
\fl \dot{n}(t) + \frac{1}{4} N \left( \gamma_x \left( \Delta,t \right) + \gamma_x \left( 2 \Omega + \Delta,t \right) + \gamma_z \left( \Omega,t \right) \right) n(t) = \nonumber \\
\fl= \frac{1}{4} N \left( \gamma_x (\Delta,t) - \gamma_x \left( 2 \Omega + \Delta,t \right) \right)   n_B + \frac{1}{4} N \left( \gamma_x \left( 2 \Omega + \Delta,t \right) + \frac{1}{2} \gamma_z \left( \Omega,t \right) \right).
\end{eqnarray}
With $ n_B$ the population of a single nuclear spin. When thermal baths are considered, $n_B = 1/2$ and we arrive to a final expression

\begin{equation}
\dot{n}(t) + \alpha(t) n(t) = \frac{1}{2} \alpha(t).
\end{equation}
With the depolarization rate $\alpha(t) \equiv  \frac{1}{4} N \left( \gamma_x ( \Delta,t ) + \gamma_x ( 2 \, \Omega + \Delta,t) + \gamma_z \left( \Omega,t \right) \right)$.

\section{Equivalence between Master Eq. and dynamical equation for covariance matrix}
\label{sec:AppendixB}
No we focus on the derivation of the evolution equation for the covariance matrix that will describe the evolution of the system when a mixture of phases is present. In fact, when the two phases coexist the description of the NV center dynamics becomes more challenging since we need to combine the evolution given by Eq.(\ref{eq:EqSolid}), with the depolarization created by a liquid bath Eq.(\ref{eq:EqLiquid}). While the former is a equation for bosons, since we have applied the HPA, the latter is a equation for spins. Since we are interested in obtain numerical results, we prefer to translate the equation obtained for moving spins to an equivalent expression for bosons.

We start tracing out the liquid bath in Eq.(\ref{eq:MasterEqSpins}) in order to obtain a master equation solely for the NV

\begin{equation}
\dot{\rho}_{NV} (t) = -{\mrm{i}} \left[ \tilde{\Omega}  S_z ,\rho_{NV} (t) \right] + \left[ \gamma_+ D \left( S_+ \right) + \gamma_- D \left( S_- \right)   \right] \rho_{NV} (t).
\end{equation}
For the sake of simplicity we have introduced the new auxiliary variables, $\tilde{\Omega}$, $\gamma_+$, $\gamma_-$, we omit its explicit form for a moment since it is not relevant for our porpoise here. Under the influence of the former master equation, the population on the NV evolves as 
\begin{equation}
\dot{n}(t) = \gamma_+ - \left( \gamma_+ + \gamma_- \right) n(t),
\end{equation}
which correspond to Eq.(\ref{eq:EvolutionEqSpinPopulation}), with $n_B = \frac{1}{2}$ and $ \gamma_+ = \gamma_- = \frac{1}{2} \alpha(t)$. If now we look at a bosonic system, we have that a master equation of the form

\begin{equation}
\label{eq:MasterEqLiquidBoson}
\dot{\rho}_{NV} (t) = -{\mrm{i}} \left[ \tilde{\Omega} \, a_0^\dagger a_0 , \rho_{NV}(t) \right] + \left[ \gamma_{a^\dagger} D \left( a_0^\dagger \right) + \gamma_a D\left(a_0 \right) \right] \rho_{NV} (t),
\end{equation}
leads to a equation for the population as

\begin{equation}
\dot{n}(t) = \gamma_{a^\dagger} + \left( \gamma_{a^\dagger} - \gamma_a \right) n(t).
\end{equation}
Therefore, the bosonic system will evolve equivalently to spin system as far as $\gamma_{a^\dagger} = \gamma_+$ and $\gamma_a = 2\, \gamma_+ + \gamma_-$. For $n_B= \frac{1}{2}$ we can again express these quantities in terms of the depolarization rate, $\gamma_{a^\dagger} = \frac{1}{2} \alpha(t)$, $\gamma_a = \frac{3}{2} \alpha(t) $. Thus, with appropriate rates, we can reproduce on a bosonic system the dynamics generated by a Lindblad dissipator on a spin system.

Provided now that for bosons we write the total Hamiltonian (NV and solid spins) as $H$, the master equation of the system is

\begin{equation}
\dot{\rho} (t) = -{\mrm{i}} \left[ H, \rho(t) \right] + \left[ \frac{1}{2} \alpha(t) D\left(a^\dagger_0\right) + \frac{3}{2} \alpha(t) D \left(a_0\right) \right] \rho(t).
\end{equation}
This master equation leads to a evolution equation for the covariance matrix of the form of Eq.(\ref{eq:CovarianceMatrixMixedWater}). On the other hand, this reflects that a direct application of the HPA on the master equation does not always lead to faithful result. In our case, this forbids us to use the HPA for obtaining results in the case of fast moving nuclei.

\section{NV magnetometry}
In this part of the Appendix we present the derivation of Eq.(\ref{eq:Envelope}). As we are principally interested in the solid part, we use the HPA and treat a system of bosons. Under the action of a high magnetic field gradient parallel to the NV center quantization axis, the Larmor frequency of a nucleus at a position $z_i$ is just $\omega_i = \gamma_H \left( B_0 + \lambda z_i \right)$, where $B_0$ is a constant magnetic field and $\lambda$ is the gradient strength $\lambda \equiv dB/dz$. The Hamiltonian for a bosonic system expressed in Eq.(\ref{eq:HamiltonianSolidWater}), can be straightforwardly modified resulting in

\begin{equation}
H= \Omega a^\dagger_0 a_0 + \sum_{i=1}^N \omega_i(z_i) a_i^\dagger a_i + \sum_{i=1}^N g_i a_0 a_i^\dagger + \frac{1}{8} \sum_{i,j=1}^N h_{i,j} a_i^\dagger a_j + \mrm{h.c.}
\end{equation}
For a crystal such as Ice Ih, the $^1$H nuclei are distributed in layers separated by certain distance $d$. Therefore, if a given layer seats at a distance $z_{L}= n \, d + z_0$ from the NV center, all the nuclei in this layer will precess with the same Larmor frequency $\omega_i \left(n d + z_0 \right)$. Moreover, they are detuned from nuclei belonging to adjacent layers, $\tilde{\Delta} = \omega_i \left( n d + z_0 \right) - \omega_j \left((n+1)d + z_0 \right) =\gamma_H \lambda d $. If our field gradient is high enough such that $\tilde{\Delta} \gg \sqrt{\sum_{z_i = z_L} \vert g_i^2 \vert}$, then setting $\Omega = \tilde{\omega} (z_L)$, the effective Hamiltonian may be written as

\begin{equation}
H = \tilde{\omega}(z_L) a^\dagger_0 a_0 + \sum_{z_i = z_L} \tilde{\omega}(z_L) a_i^\dagger a_i + \sum_{z_i=z_L} g_i a_0 a_i^\dagger +\frac{1}{8} \sum_{i,j=1}^N h_{i,j} a_i^\dagger a_j + \mrm{h.c.}
\end{equation}
Note that the sum expands only in a given layer. In the limit of short times, and $g^i > h_{i,j}$, we can neglect the internuclear coupling obtaining

\begin{equation}
H \approx \tilde{\omega}(z_L) a^\dagger_0 a_0 + \sum_{z_i = z_L} \tilde{\omega}(z_L) a_i^\dagger a_i + \sum_{z_i=z_L} g_i a_0 a_i^\dagger + \mrm{h.c} .
\end{equation}
The latter Hamiltonian may be solved analytically. When at the initial time the NV is polarized while the bosonic bath is in thermal state, we obtain

\begin{equation}
n ( \Omega = \tilde{\omega} (z_L),t ) = \frac{1}{2} + \frac{1}{2} \cos^2 \left( \sqrt{\sum_{z_i=z_L} \vert g_i \vert^2 } t  \right).
\end{equation}
Recalling now the expression for the coupling strength $g_i = \frac{1}{4} \left( A_x^i + \mrm{i} A_y^i \right)$, we can compute the sum as

\begin{equation}
\fl \sum_{z_i=z_L} \vert g_i ^2 \vert = \frac{1}{4} \sum_{z_i = z_L} \vert {A_x^i} \vert^2 + \vert {A_y^i} \vert^2 \propto \frac{1}{4} \sum_{z_i = z_L} \frac{\left( x_i^2 +y_i^2 \right) z_i^2 }{\left( x_i^2 + y_i^2 + z_i^2 \right)^{5}} \approx \frac{1}{4} \rho_{2D} \int_V \frac{\left( x^2 +y^2 \right) z_L^2 }{\left( x^2 + y^2 + z_L^2 \right)^{5}} dx dy.
\end{equation}
In the last expression we have introduced the bidimensional density of protons $\rho_{2D}$. The expression for the population is

\begin{equation}
\label{eq:PopulationatPeaks}
n( \Omega = \tilde{\omega} (z_L),t )= \frac{1}{2} + \frac{1}{2} \cos^2 \left( \sqrt{ \rho_{2D}^i} \frac{\beta}{z_L^2} t \right).
\end{equation}
which corresponds to Eq.(\ref{eq:Envelope}).

\section*{References}
\bibliographystyle{unsrt}

\end{document}